\pdfoutput=1

\documentclass[a4paper]{article}
\usepackage{textcomp}
\usepackage[tight]{units}
\usepackage{amsmath,color}
\usepackage{amsfonts}
\usepackage{fixmath}
\usepackage{braket}

\usepackage{natbib}

\usepackage{graphicx}
\usepackage{float}
\usepackage{fixmath}
\usepackage{epstopdf}
\usepackage{verbatim}
\usepackage{subfig}
\usepackage{hyperref}

\usepackage{mathtools}

\newcommand{\bx}{\mathbf{x}}

\title{Study of cost functionals for ptychographic phase retrieval to improve the robustness against noise, and a proposal for another noise-robust ptychographic phase retrieval scheme}

\author{A.P. Konijnenberg$^1$, W.M.J. Coene$^{1,2}$, H.P. Urbach$^1$}

\begin{document}
\maketitle
\noindent
$^1$Optics Research Group, Delft University of Technology, Delft 2628 CH, Netherlands\\
$^2$ ASML Netherlands B.V., De Run 6501, 5504 DR Veldhoven, The Netherlands
\\
\\
\centerline
{
a.p.konijnenberg@tudelft.nl
}

\begin{abstract}
Recently, efforts have been made to improve ptychography phase retrieval algorithms so that they are more robust against noise. Often the algorithm is adapted by changing the cost functional that needs to be minimized. In particular, it has been suggested that the cost functional should be obtained using a maximum-likelihood approach that takes the noise statistics into account. Here, we consider the different choices of cost functional, and to how they affect the reconstruction results. We find that seemingly the only consistently reliable way to improve reconstruction results in the presence of noise is to reduce the step size of the update function. In addition, a noise-robust ptychographic reconstruction method has been proposed that relies on adapting the intensity constraints
\end{abstract}

\section{Introduction}

The problem of phase retrieval occurs in many different fields, such as Coherent Diffractive Imaging (CDI), microscopy, pulse reconstruction, and quantum information. Mathematically, we can describe the problem as follows: we have a complex-valued field $f(x)$ with Fourier transform (i.e. far field) $F(u)$
\begin{equation}
F(u)=\sum_{x'} f(x')e^{-2\pi i u x'}\Delta_x.
\end{equation}
In the context of imaging, $x$ is a 2D position vector in object space (in the case of CDI), and $u$ is a 2D vector in Fourier space. We cannot measure complex fields directly, but only amplitudes $|F(u)|$ can be known from measurement (or technically the intensity $|F(u)|^2$). Given these amplitudes amplitude measurements in Fourier, one uses additional constraints in object space to reconstruct the complex fields. In the case of the Gerchberg-Saxton algorithm \cite{gerchberg1972practical}, this constraint is the amplitude in object space $|f(x)|$. In case of single-measurement algorithms such as the Error-Reduction (ER) algorithm or Hybrid Input-Output (HIO) algorithm \cite{fienup1978reconstruction}, the object-space constraint is a support constraint, i.e. we know that $f(x)=0$ for $x\notin S$, where $S$ is the known support of $f(x)$. In the case of ptychography \cite{rodenburg2004phase}, we have multiple diffraction patterns $|F_j(u)|$ corresponding to different fields $f_j(x)$ in object space, which are obtained by scanning a thin object $O(x)$ with a probe $P(x)$ over different probe positions $X_j$
\begin{equation}
f_k(x)=O(x)P(x-X_j).
\end{equation}
The probe positions $X_j$ are chosen such that the probes at different positions overlap with each other, giving in the object space an overlap constraint. Phase retrieval with ptychography has recently gained interest due to its robustness against noise and aberrations in the probe \cite{maiden2009improved}.
\\
\\
There are several ways to formulate the phase retrieval problem, leading to different approaches to solve them. Let us consider two important perspectives:
\begin{itemize}
\item \textbf{Finding the intersection of two sets}: we have a set $M$ of all functions $g(x)$ which have a Fourier transform that satisfy the Fourier constraint, and we have a set $S$ of all $g(x)$ that satisfy the object constraint. For example, for a single-measurement phase retrieval problem, we have in Fourier space the modulus constraint $|G(u)|=|F(u)|$
\begin{equation}\label{modconst}
M=\{g(x): |G(u)|=|F(u)|\},
\end{equation}
and we the support constraint in object space $g(x)\mathbf{1}_{x\notin S} = 0$
\begin{equation}
S=\{g(x): g(x)\mathbf{1}_{x\notin S} = 0\}.
\end{equation}
The phase retrieval problem can then be formulated as follows \cite{bauschke2002phase}: find some $g(x)\in S\cap M$. From this point of view, the problem is usually approached with alternate projection methods, which alternately project $g(x)$ onto $M$ and onto $S$. This approach gives insights into algorithms such as ER, HIO, RAAR \cite{luke2004relaxed} or HPR \cite{bauschke2003hybrid}.
\item \textbf{Minimizing a cost function}: we have a cost functional $L[g(x)]$ that quantifies how much a function $g(x)$ (that obeys the object space constraint) satisfies the Fourier space constraint. The problem we have to solve is the following: find $g(x)$ such that $L[g(x)]$ is minimized. The most basic method to solve this is the Steepest Descent method (which for single-measurement phase retrieval leads to the ER algorithm \cite{Fienup1982}, and which for ptychography leads to an algorithm similar to the Ptychographical Iterative Engine (PIE) \cite{GuizarSicairos2008}), although there are more sophisticated (higher-order) methods such as the Newton Raphson method, or the Conjugate Gradient method \cite{qian2014efficient}.
\end{itemize}
These different methods may try to achieve different goals: for example, they aim to avoid stagnation issues, or try to reduce the required computational power by making the algorithm converge in fewer iterations. In this paper we focus on making the algorithm more robust against noise. Several methods from the literature are explained and considered more in depth, and new options are considered as well.

\section{Deriving the update function from a cost functional}
A phase retrieval algorithm can be derived by minimizing a cost functional $L(g_k(x))$ with a Steepest Descent scheme
\begin{equation}\label{descent}
g_{k+1}(x)=g_k(x)-\mu \frac{\text{d}L}{\text{d}g_k(x)^*},
\end{equation}
where $\frac{\text{d}L}{\text{d}g_k(x)^*}$ is a Wirtinger derivative that indicates the direction of steepest ascent, and $\mu$ is the step size. To illustrate this, let us see how the Error Reduction algorithm is derived from an amplitude-based cost function as first demonstrated by \cite{Fienup1982}. Here, $g_k(x)$ is the $k^{\text{th}}$ estimate of $f(x)$. We know that $f(x)$ has a finite support $S$, i.e. $f(x)=0$ if $x\notin S$. 
We define the cost function that is to be minimized
\begin{equation}\label{cost1}
L[g_k(x)]=\sum_u (|G(u)|-|F(u)|)^2,
\end{equation}
where 
\begin{equation}
\begin{split}
G(u)&=\sum_{x'} \mathbf{1}_{x\in S}g(x')e^{-2\pi i u x'}\Delta_x,
\\
F(u)&=\sum_{x'} f(x')e^{-2\pi i u x'}\Delta_x.
\end{split}
\end{equation}
We can calculate
\begin{equation}\label{derivL1}
\begin{split}
\frac{\text{d}L}{\text{d}g_k^*(x)}&=\frac{\text{d}L}{\text{d}|G(u)|}\frac{\text{d}\sqrt{|G(u)|^2}}{\text{d}|G(u)|^2}\frac{\text{d}|G(u)|^2}{\text{d}G(u)^*}\frac{\text{d}G(u)^*}{\text{d}g_k(x)^*}
\\
&=\sum_u (|G(u)|-|F(u)|)\frac{G(u)}{|G(u)|}e^{2\pi i u x}\Delta_x\mathbf{1}_{x\in S}
\\
&=\left(g_k(x)-\mathcal{F}^{-1}\left\{\frac{G(u)}{|G(u)|}|F(u)|\right\}\right)\mathbf{1}_{x\in S}.
\end{split}
\end{equation}
We define
\begin{equation}\label{gprime}
g_k'(x)=\mathcal{F}^{-1}\left\{\frac{G(u)}{|G(u)|}|F(u)|\right\}.
\end{equation}
Note that $g_k'(x)$ is what we get when we take the Fourier transform of $g_k(x)$, substitute its amplitude with the measured amplitude $|F(u)|$ with keeping the phase of $G(u)$, and inverse Fourier transforming the result. When we substitute the outcome of Eq. \eqref{derivL1} in Eq. \eqref{descent}, we obtain the iterative update scheme for $g_k(x)$
\begin{equation}\label{update1}
\begin{split}
O_{k+1}(x)&=g_{k}(x)-\mu\left(g_k(x)-g_k'(x)\right)\mathbf{1}_{x\in S}
\\
&=\begin{cases}
(1-\mu)g_k(x) + \mu g_k'(x) &\quad \text{if $x\in S$}
\\
0 &\quad \text{if $x\notin S$,}
\end{cases}
\end{split}
\end{equation}
where we assumed that $g_k(x)\mathbf{1}_{x\notin S}=0$. In particular, for $\mu=1$ this reduces to
\begin{equation}\label{errorreduction}
\begin{split}
g_{k+1}(x)&=g_k'(x) \mathbf{1}_{x\in S}
\\
&=\begin{cases}
g_k'(x) &\quad \text{if $x\in S$}
\\
0 &\quad \text{if $x\notin S$.}
\end{cases}
\end{split}
\end{equation}
This is the Error Reduction scheme, which can also be derived using the method of alternate projections.

\subsection{Ptychography}
In the case of ptychography, we define $f(x,X)$ to be the exit wave for probe position $X$
\begin{equation}
f(x,X)=P(x-X)O(x),
\end{equation}
and $g_k(x,X)$ the $k^{\text{th}}$ estimate of the exit wave for probe position $X$
\begin{equation}
g_k(x,X)=P(x-X)O_k(x),
\end{equation}
where $O_k(x)$ is the $k^{\text{th}}$ estimate of the object. We can define the update function by applying sequentially for each probe position
\begin{equation}\label{ptychupdate}
O_{k+1}(x)=O_k(x)+\mu P(x-X)^*(g_k'(x,X)-g_k(x,X)),
\end{equation}
where $g_k'(x,X)$ is defined as in Eq. \eqref{gprime}. However, note that if we define the cost function 
\begin{equation}
L[O(x)]=\sum_X \sum_u (|G(u,X)|-|F(u,X)|)^2,
\end{equation}
then according to the steepest descent scheme we should not apply the update to each position sequentially, but rather apply an update using the information of all probe positions simultaneously 
\begin{equation}
\begin{split}
\frac{\text{d}L}{\text{d}O_k(x)^*}&=\sum_X \sum_u \frac{\text{d}(|G(u,X)|-|F(u,X)|)^2}{\text{d}g_k(x,X)^*}\frac{\text{d}g_k(x,X)^*}{\text{d}O(x)^*}
\\
&=\sum_X (g_k(x,X)-g_k'(x,X))P(x-X)^*.
\end{split}
\end{equation}
It turns out though that applying updates sequentially gives better results \cite{Konijnenberg2016}, so we will stick to this approach.

\section{Different cost functionals}
To derive the Error Reduction update function, we chose an amplitude-based cost functional as in Eq. \eqref{cost1}. One might wonder what is so special about this amplitude-based cost functional. Why not for example choose an intensity-based cost functional
\begin{equation}\label{cost1}
L[g_k(x)]=\sum_u \left(|G(u)|^2-|F(u)|^2\right)^2,
\end{equation}
or some other cost functional? In the absence of noise the precise choice of the functional does not affect the reconstruction that much, since the minimum is still located at $|G(u)|=|
F(u)|$. However, when the measurements $|F(u)|$ are corrupted by noise, then $|G(u)|=|
F(u)|$ is in general not an allowed solution because of the object-space constraint (either a support constraint in the case of a single intensity measurement, or the constraint that the exit waves can be factorized in an object and shifted probe in the case of ptychography), and the minimum for $L$ must lie elsewhere. In this case, the precise form of the cost functional might indeed matter, as was also found by \cite{Yeh2015}. In the next few sections we discuss different forms of cost functionals which may improve the robustness of the algorithms against noise.

\subsection{The log-likelihood cost functional}
To explain the motivation for a log-likelihood cost functional, let us introduce the following notation:
\begin{itemize}
\item $f(x)$ is the to be reconstructed complex-valued function, and $F(u)$ is its Fourier transform.
\item $g_k(x)$ is the $k^{\text{th}}$ estimate of $f(x)$, and it has Fourier transform $G_k(u)$.
\item $m$ denotes what would be the measured intensity in the noise-free case
\begin{equation}
m=|F(u)|^2.
\end{equation} 
\item $z$ denotes the estimated intensity
\begin{equation}
z=|G_k(u)|^2.
\end{equation}
\item $y$ denotes the actually measured intensity when noise is present. For example, when the intensity measurement is affected by Poisson noise, then
\begin{equation}
y \sim \text{Poisson}(m).
\end{equation}
\end{itemize}
With this notation, the intensity-based cost function reads
\begin{equation}\label{intcost}
L=\sum_{u}(z-y)^2,
\end{equation}
and the amplitude-based cost function reads
\begin{equation}
L=\sum_{u}(\sqrt{z}-\sqrt{y})^2.
\end{equation}
We will explain two different viewpoints from which the log-likelihood cost can be motivated: from the maximum-likelihood principle, a from a variance-stabilization principle.
\\
\\
Let us derive the log-likelihood cost function from a maximum-likelihood principle \cite{Thibault2012,Bian2016}. In this method, we assume a certain noise model: given the noise-free value $m$, we know the probability $P(y|m)$ that we measure a noisy value $y$. The probability of measuring a set of noisy values $y(u)$ given a set of noise-free values $m(u)$ is then given by
\begin{equation}
P_{\text{tot}}[m(u)]=\prod_u P(y|m).
\end{equation} 
In the maximum-likelihood method we try to find the $g(x)$ for which the set of measurements $y(u)$ is most likely, i.e. we try to find $g(x)$ for which $P_{\text{tot}}[z(u)]$ is maximized. Equivalently, we can try to find $g(x)$ for which $-\log P_{\text{tot}}[z(u)]$ is minimized. Thus, we can define the log-likelihood cost function as follows
\begin{equation}\label{costlog}
L[g(x)]=-\sum_u \log P(y|z).
\end{equation}
For a concrete example, we can consider shot noise in which case $P(y|m)$ follows a Poisson distribution
\begin{equation}
P(y|m)=\frac{m^y e^{-m}}{y!},
\end{equation}
in which case
\begin{equation}\label{loglik}
L[g(x)]=\sum_u z-y\log(z),
\end{equation}
where we neglect the term $\log(y!)$ because this term does not vary with $g(x)$ and therefore is irrelevant for the minimization problem. We can calculate
\begin{equation}
\frac{\text{d}L}{\text{d}g_k^*(x)}=2\left(g_k(x)-\mathcal{F}^{-1}\left\{\frac{G}{|G|^2}y\right\}\right),
\end{equation}
which we can plug in Eq. \eqref{descent} to find an update formula. Note that this update formula can be problematic when $|G|$ is small. Therefore, a very good initial guess is required, or some regularization to prevent divergence is required \cite{Thibault2012,Bian2016}.
\\
\\
We have derived the log-likelihood cost functional from a maximum likelihood principle. Let us view the same cost functional from a variance-stabilization perspective. When considering Poisson noise, the problem with an intensity-based cost functional as in Eq. \eqref{intcost} is the following: for each pixel $u$, the difference between the measured value and the estimated value $(z(u)-y(u))^2$ is being minimized. In the noise-free case, there is no problem because $z(u)=y(u)$ is a valid solution. However, in the presence of noise the expected value of $\Delta(u)=(m(u)-y(u))^2$ is non-zero, and more importantly, the expected value of $\Delta(u)$ differs per pixel $u$. More specifically, for Poisson noise $\Delta(u)$ is expected to be larger for larger values of $m(u)$, while the intensity-based cost functional of Eq. \eqref{intcost} does not take this into account: it tries to minimize all $(z(u)-y(u))^2$ equally for all $u$, regardless of $y(u)$ or $m(u)$.
\\
\\
So let us try to find for each $u$ an expression that needs to be minimized equally for all pixels $u$, regardless of their measured intensity $y(u)$ or noise-free intensity $m(u)$. The probability that given a noise-free intensity $m$ we measure $y$ is given by $P(y|m)$. Conversely, given a measured value $y$, the probability that the noise-free value (which we try to reconstruct with $z$) is $m$ is $P_y(m)=P(y|m)$. We see that the probability distribution of $m$ depends on the measured value $y$. We want to find a transformation $T_y$ that transforms $m$ to a variable $m'$, the probability distribution $P'(m')$ of which is the same for all pixels. Let us say we want $m'$ to be normally distributed with mean 0 and standard deviation 1
\begin{equation}
\begin{split}
T_y(m)&: m \to m'
\\
m' &\sim \mathcal{N}(0,1),
\\
P'(m')&=\frac{e^{-m'^2/2}}{\sqrt{2\pi}}.
\end{split}
\end{equation}
We can define the cost function we want to minimize
\begin{equation}\label{costlog2}
L[g_k(x)]=\sum_u z'(u)^2,
\end{equation}
where $z'=T_y(z)$. To find how to transform $m\to m'$, we thus need to solve
\begin{equation}
\begin{split}
\alpha(y) P_y(m) &= e^{-m'^2/2} 
\\
&\propto P'(m'),
\end{split}
\end{equation}
where $\alpha(y)$ takes care of any required normalizations
\begin{equation}
\alpha(y) = \frac{1}{\underset{m}{\text{max}}\, P_y(m)}.
\end{equation}
Solving this gives
\begin{equation}
m'^2= -2\log \left(P_y(m) \right) - 2\log(\alpha).
\end{equation}
Using this result in Eq. \eqref{costlog2}, we find up to an irrelevant additive and multiplicative constant the log-likelihood cost functional as in Eq. \eqref{costlog}. Thus we have seen that the log-likelihood functional can be interpreted in two different ways:
\begin{itemize}
\item It is a cost functional which, when minimized, gives the object for which the measured intensity patterns would be most likely.
\item It is a cost functional for which each term (i.e. for each pixel $u$) obeys the same probability distribution, and thus should be minimized equally.
\end{itemize}
In the next section, we see how the second viewpoint can lead to other forms of cost functionals which suffer less from divergence problems the log-likelihood cost functional suffers from.

\subsection{Variance stabilizing transforms}
We have seen that the log-likelihood cost function, which in the literature has been derived from a maximum-likelihood principle, can be interpreted as a method to make each term in the cost functional obey the same probability distribution. Another method to do this is by using variance stabilizing transforms \cite{Godard2012,Marchesini2016,Zhang2017}. For example, if $y \sim \text{Poisson} (m)$, then the variance of $y$ is directly proportional to $m$. However, the variance of $\sqrt{y}$ is more or less independent of $m$ (especially for larger values of $m$). Thus, for a Poisson distributed variable $y$ with mean $m$, the transformation $T(y)=\sqrt{y}$ is a variance-stabilizing transform, and $T(y)-T(m)$ has mean 0 and variance independent of $m$. Thus, one can introduce the cost functional
\begin{equation}
L[g_k(x)]=\sum_u (T(y)-T(z))^2,
\end{equation}
where $T$ is a variance stabilizing transform, and where for $T(y)=\sqrt{y}$ one obtains the amplitude-based cost function of Eq. \eqref{cost1}. Also note that this cost functional can by obtained by Taylor expanding the terms in the log-likelihood functional of Eq. \eqref{loglik} in terms of $\sqrt{z}$ around the point $\sqrt{z}=\sqrt{y}$
\begin{equation}
z-y\log z \approx y-y\log y + 2(\sqrt{z}-\sqrt{y})^2,
\end{equation} 
where the additive constant is irrelevant for the cost functional. Thus, we can motivate the use of a variance-stabilized cost functional by noting that the solution of the minimization problem depends only on the behaviour of the cost functional around its minimum $z(u)=y(u)$. In order to avoid problems of divergence as in the case of a log-likelihood functional, it is thus acceptable to instead use a variance stabilizing transform, or to Taylor expand the log-likelihood functional (which in this case gives similar results). It has been known however that for a Poisson distributed variable, the Anscombe $T(y)=\sqrt{y+3/8}$ is a better variance stabilizing transform than $T(y)=\sqrt{y}$ \cite{Anscombe1948}. Thus, if the aforementioned reasoning holds water, using a cost functional that uses the Anscombe transform should give better reconstruction results than the standard amplitude-based cost functional, provided that the noise is Poisson distributed. For the case of Fourier ptychography, this has been tested by \cite{Zhang2017}.
 
\subsection{Reducing the step size of the update function}
In Eq. \eqref{descent} it is described how one can derive an update scheme from a cost functional, after which we proceeded to discuss several choices for cost functionals. However, we have not yet considered the influence of the step size $\mu$ on the reconstruction. To prevent the algorithm from overshooting around the minimum of $L$, it seems sensible to decrease the step size $\mu$ in the final iterations (although other sources suggest increasing the step size \cite{Bian2016}). Thus, decreasing $\mu$ is expected to improve the reconstruction quality, as has been demonstrated and explained in more detail by \cite{Zuo2016}. We have seen in Eq. \eqref{errorreduction} that for $\mu=1$ the steepest-descent algorithm for phase retrieval with a single intensity measurement coincides with the Error Reduction algorithm, which can also be derived using alternate projections. Suppose we change $\mu$, what would that mean in the context of projections, and what variation of the algorithm can be designed from this point of view?
\\
\\
Let us rewrite the ptychographical update function from Eq. \eqref{ptychupdate} as
\begin{equation}
O_{k+1}(x)=O_{k}(x)+P(x-X)^* \left(\mathcal{F}^{-1}\left\{\left(\mu|F(u)|+(1-\mu)|G(u)|\right)\frac{G(u)}{|G(u)|}\right\}-g_k(x)\right).
\end{equation}
We can see that for $\mu=1$, the Fourier modulus $|G(u)|$ of the exit wave $g_k(x)$ is replaced with the measured (in this example noise-free) amplitude $|F(u)|$ while keeping the phase $\frac{G(u)}{|G(u)|}$ of estimated diffraction field. In other words, we have projected $G(u)$ into $M$ as defined in Eq. \eqref{modconst}. If we choose $\mu$ smaller than 1, then we substitute $|G(u)|$ with the convex combination $\mu|F(u)|+(1-\mu)|G(u)|$, i.e. rather than setting the amplitude to $|F(u)|$, we `move from $|G(u)|$ towards $|F(u)|$' while keeping the phase $\frac{G(u)}{|G(u)|}$ the same. In \cite{Soulez2016} another method is proposed to not completely go from $|G(u)|$ to $|F(u)|$.
\\
\\
Instead of using a convex combination of the amplitudes $|G(u)|$ and $|F(u)|$
\begin{equation}\label{convexcombination1}
G(u) \to \left[(1-\mu)|G(u)|+\mu|F(u)|\right] \frac{G(u)}{|G(u)|},
\end{equation}
one may wonder why not to use a convex combination of e.g. the intensities
\begin{equation}
G(u) \to \sqrt{(1-\mu)|G(u)|^2+\mu|F(u)|^2} \frac{G(u)}{|G(u)|},
\end{equation}
or more generally, for some transformation $T(z)$
\begin{equation}\label{Tz1}
G(u) \to \sqrt{T^{-1}\left((1-\mu)T(|G(u)|^2)+\mu T(|F(u)|^2)\right)} \frac{G(u)}{|G(u)|}.
\end{equation}
Again it may be argued that a variance stabilizing transform of the form $T(z)=\sqrt{z+\alpha}$ may improve the performance of the algorithm in the presence of noise.
\\
\\
Alternatively, defining
\begin{equation}
O'_k(x)=O_k(x)+P(x-X)^*(g_k'(x)-g_k(x))
\end{equation}
we can rewrite the ptychographical update function from Eq. \eqref{ptychupdate} as
\begin{equation}\label{convexcombination2}
O_{k+1}(x)=(1-\mu) O_k(x) + \mu O'_k(x).
\end{equation}
Thus, whereas in Eq. \eqref{convexcombination1} the update is applied by calculating a convex combination in diffraction space, in Eq. \eqref{convexcombination2} the update is applied by calculating a convex combination in the object space. Again, one may introduce a transformation $T(z)$ and apply the update
\begin{equation}\label{Tz2}
O_{k+1}(x)=T^{-1}\left((1-\mu) T(O_k(x)) + \mu T(O'_k(x))\right).
\end{equation}
When choosing $T(z)=z^{\alpha}$ with $\alpha\in (0,1)$, it means that large differences between $O_k(x)$ and $O_k'(x)$ affect the update $O_{k+1}(x)$ more strongly than small differences.

\section{Testing different update schemes}
A large number of different adaptions have been suggested in the literature and a few have been added here. In order to test whether these algorithms work and whether the reasonings behind them are valid, the following needs to be taken into account: if the algorithm is improved just by taking into account the noise statistics (as is the case for the maximum likelihood method and variance stabilizing methods), then there should be an improvement of the reconstruction 
\begin{itemize}
\item regardless of the object $O(x)$ one tries to reconstruct,
\item regardless of the mode of ptychography (standard ptychography or Fourier ptychography \cite{Zheng2013}) that is being used,
\item if for a different noise model (e.g. exponentially distributed noise, as is the case for speckle noise \cite{Goodman1976}) the same reasoning is applied.
\end{itemize}
We test the following update schemes:
\begin{enumerate}
\item Eq. \eqref{descent}, $L=\sum_{u}(\sqrt{z}-\sqrt{y})^2$ with step size $\mu=1$. This gives the regular Error Reduction (i.e. alternate projection scheme).
\item 100 iterations of Error Reduction, then Eq. \eqref{descent}, $L=\sum_{u}(\sqrt{z}-\sqrt{y})^2$ with step size $\mu=0.1$.
\item 100 iterations of Error Reduction, then Eq. \eqref{descent}, $L=\sum_{u}(z^{0.7}-y^{0.7})^2$ with step size $\mu=0.1$.
\item 100 iterations of Error Reduction, then Eq. \eqref{descent}, $L=\sum_{u}(z^{0.9}-y^{0.9})^2$ with step size $\mu=0.1$.
\item 100 iterations of Error Reduction, then Eq. \eqref{descent}, $L=\sum_{u}(\sqrt{z+3/8}-\sqrt{y+3/8})^2$ with step size $\mu=0.1$.
\item 100 iterations of Error Reduction, then Eq. \eqref{descent}, $L=\sum_{u}(\sqrt{z+1}-\sqrt{y+1})^2$ with step size $\mu=0.1$.
\item 100 iterations of Error Reduction, then Eq. \eqref{descent}, $L=\sum_{u}(\log{(z+1/2)}-\log{(y+1/2)})^2$ with step size $\mu=0.1$.
\item 100 iterations of Error Reduction, then Eq. \eqref{descent}, $L=\sum_{u}(\log{(z+1)}-\log{(y+1)})^2$ with step size $\mu=0.1$.
\item 100 iterations of Error Reduction, then Eq. \eqref{Tz1}, $T(z)=\sqrt{z+3/8}$ with step size $\mu=0.1$.
\item 100 iterations of Error Reduction, then Eq. \eqref{Tz1}, $T(z)=\sqrt{z+1}$ with step size $\mu=0.1$.
\item 100 iterations of Error Reduction, then Eq. \eqref{Tz1}, $T(z)=z^{0.7}$ with step size $\mu=0.1$.
\item 100 iterations of Error Reduction, then Eq. \eqref{Tz1}, $T(z)=z$ with step size $\mu=0.1$.
\item 100 iterations of Error Reduction, then Eq. \eqref{Tz1}, $T(z)=\log{(z+1/2)}$ with step size $\mu=0.1$.
\item 100 iterations of Error Reduction, then Eq. \eqref{Tz1}, $T(z)=\log{(z+1)}$ with step size $\mu=0.1$.
\item 100 iterations of Error Reduction, then Eq. \eqref{Tz2}, $T(z)=\sqrt{z+3/8}$ with step size $\mu=0.1$.
\item 100 iterations of Error Reduction, then Eq. \eqref{Tz2}, $T(z)=\sqrt{z+1}$ with step size $\mu=0.1$.
\item 100 iterations of Error Reduction, then Eq. \eqref{Tz2}, $T(z)=z^{0.7}$ with step size $\mu=0.1$.
\item 100 iterations of Error Reduction, then Eq. \eqref{Tz2}, $T(z)=z$ with step size $\mu=0.1$.
\item 100 iterations of Error Reduction, then Eq. \eqref{Tz2}, $T(z)=\log{(z+1/2)}$ with step size $\mu=0.1$.
\item 100 iterations of Error Reduction, then Eq. \eqref{Tz2}, $T(z)=\log{(z+1)}$ with step size $\mu=0.1$.
\end{enumerate}
The results are shown in Figs. \ref{fig:resultsA} and \ref{fig:resultsB}. What we see is the following:
\begin{itemize}
\item Reducing the step size of the update is a reliable way to reduce the error of the reconstruction in the presence of noise.
\item Whether or not a certain choice of update function improves the reconstruction depends strongly on the object we try to reconstruct, and which mode of ptychography (real space or Fourier space ptychography) we use. This seems to contradict the idea that the noise statistics determine the optimal update function. It is worth noting that the difference between real space ptychography and Fourier space ptychography essentially comes down to a different choice of object: Fourier ptychography with an object $O(\bx)$ is the same as real space ptychography with an object $\mathcal{F}\{O(\bx)\}(\bx')$.
\item For Fourier space ptychography choosing an alternative update function seems much more beneficial than for real space ptychography, which is consistent with the findings of \cite{Zhang2017}. However, it has to be checked per type of object which update function is most beneficial.
\end{itemize}

\begin{figure}[H]
\centering
\subfloat{\includegraphics[width=0.99\textwidth]{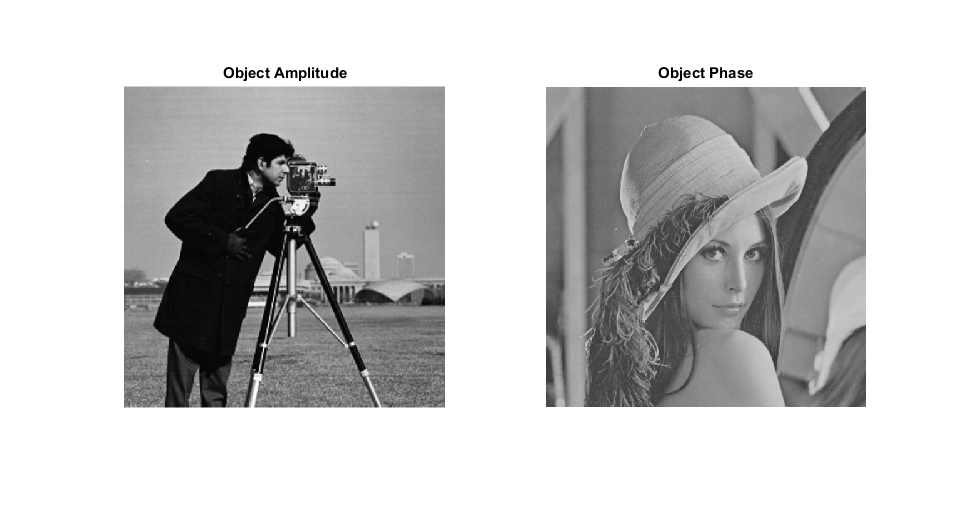}\label{fig:Lena}}
\\
\subfloat{\includegraphics[width=0.99\textwidth]{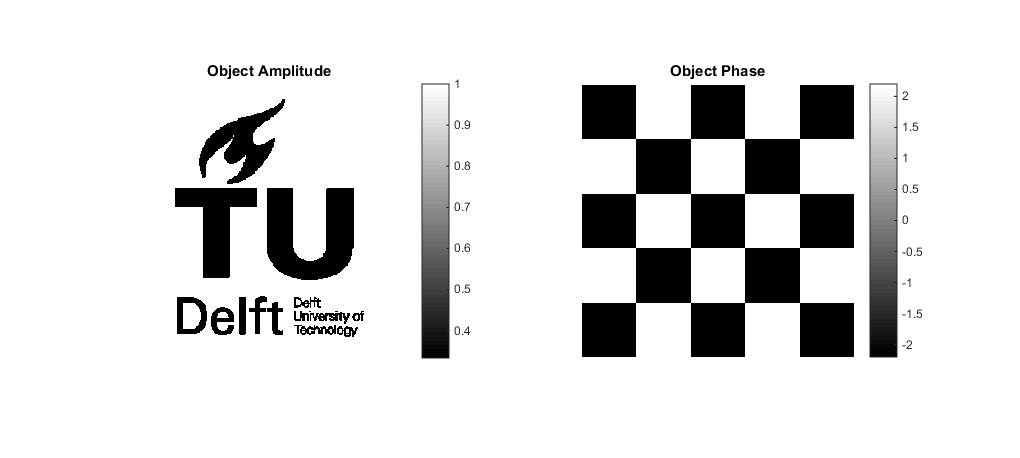}\label{fig:Delft}}
		\caption{Objects used to test the reconstruction algorithms}
		\label{fig:objects}
\end{figure}

\begin{figure}[H]
\centering
\subfloat[Fourier ptychography, Poisson noise]{\includegraphics[width=0.69\textwidth]{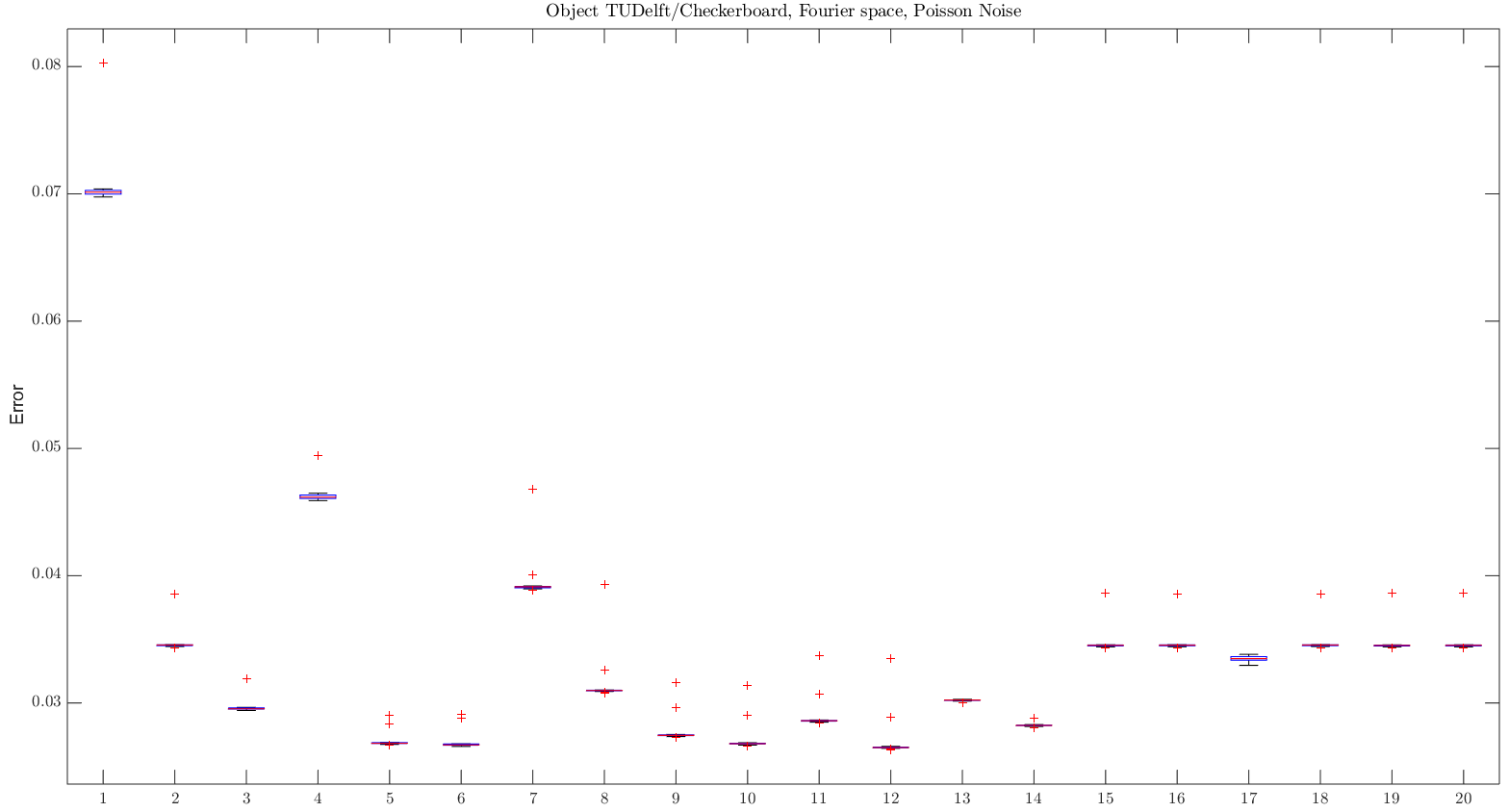}}
\\
\subfloat[Fourier ptychography, Speckle noise]{\includegraphics[width=0.69\textwidth]{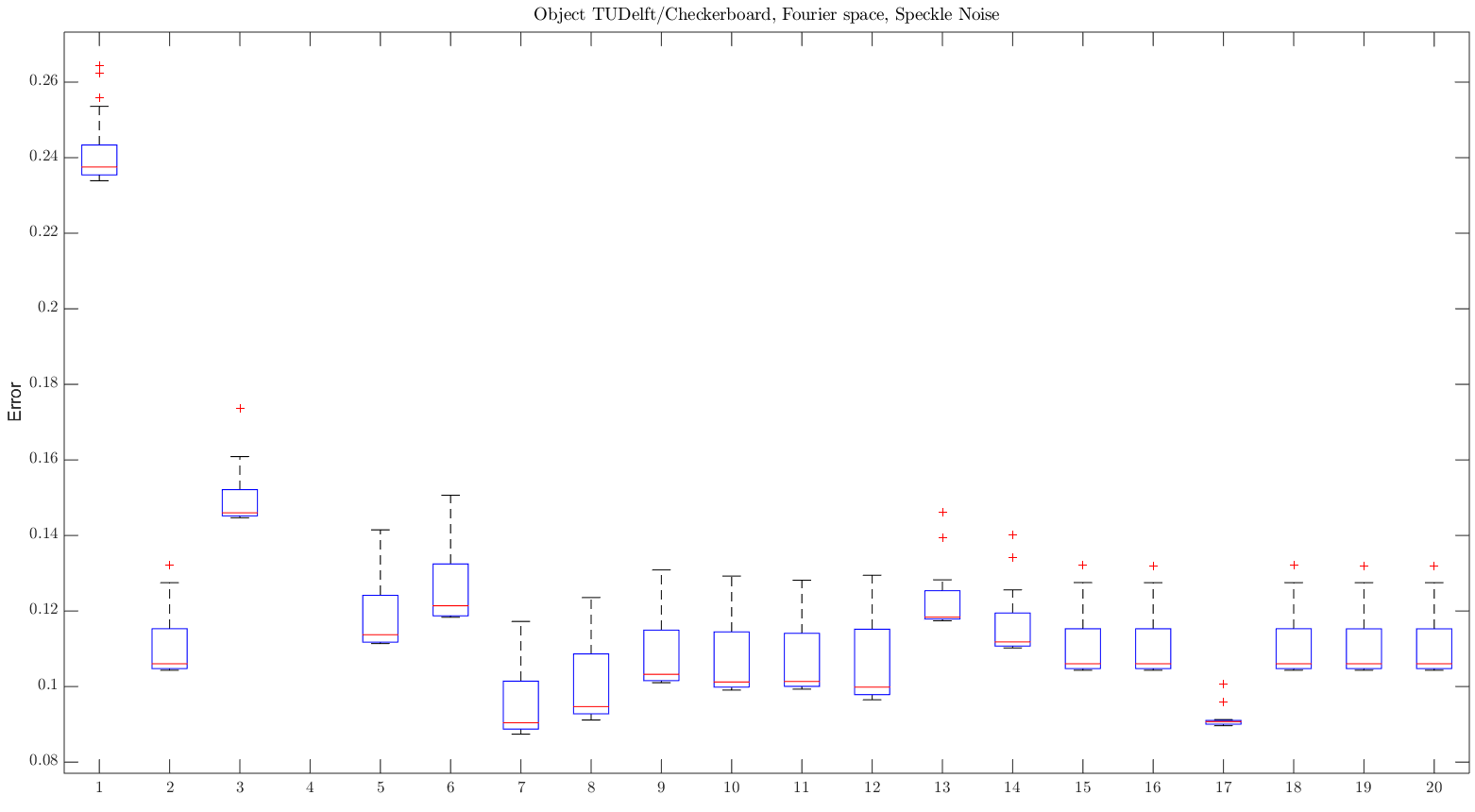}}
\\
\subfloat[Real space ptychography, Poisson noise]{\includegraphics[width=0.69\textwidth]{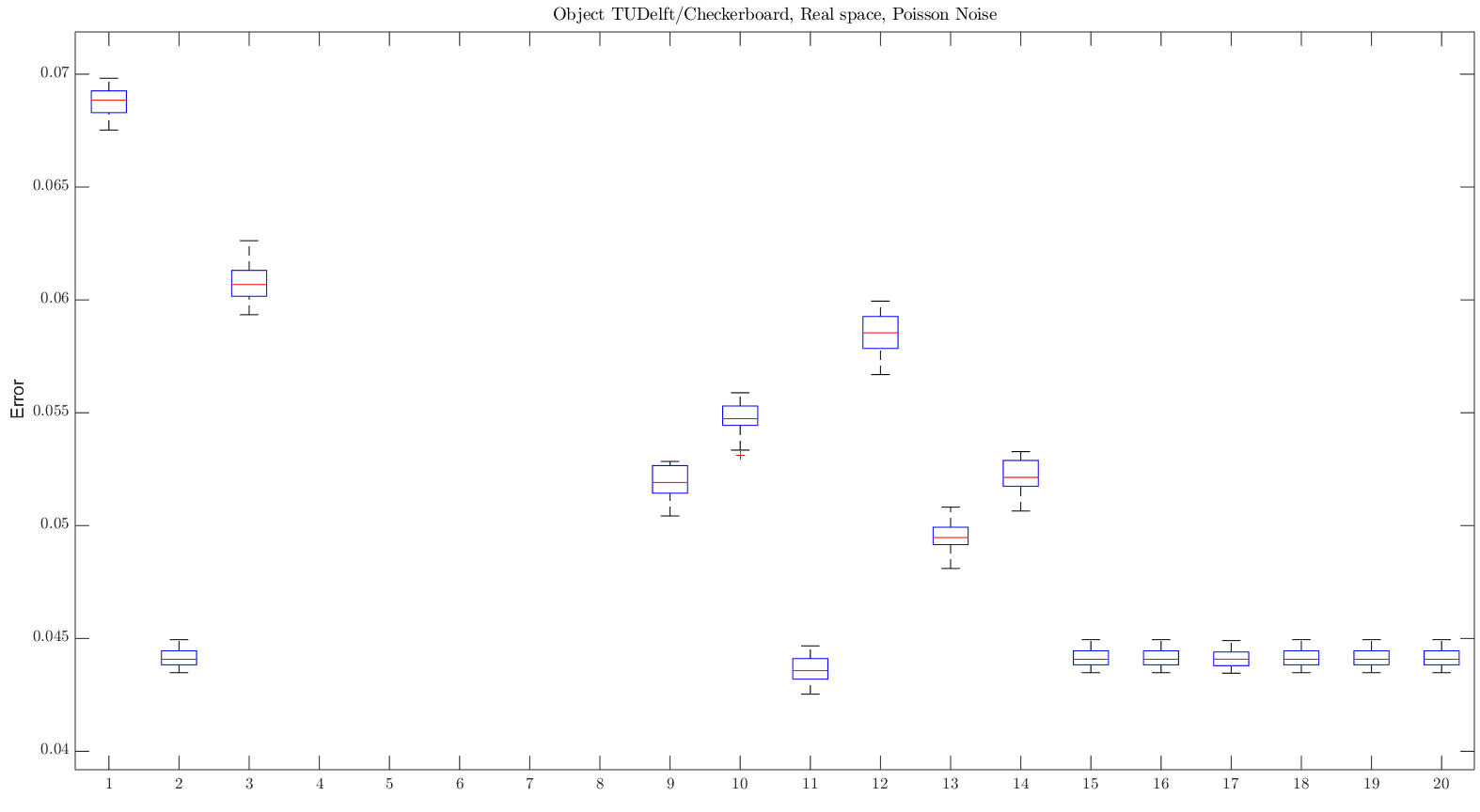}}
\\
\subfloat[Real space ptychography, Speckle noise]{\includegraphics[width=0.69\textwidth]{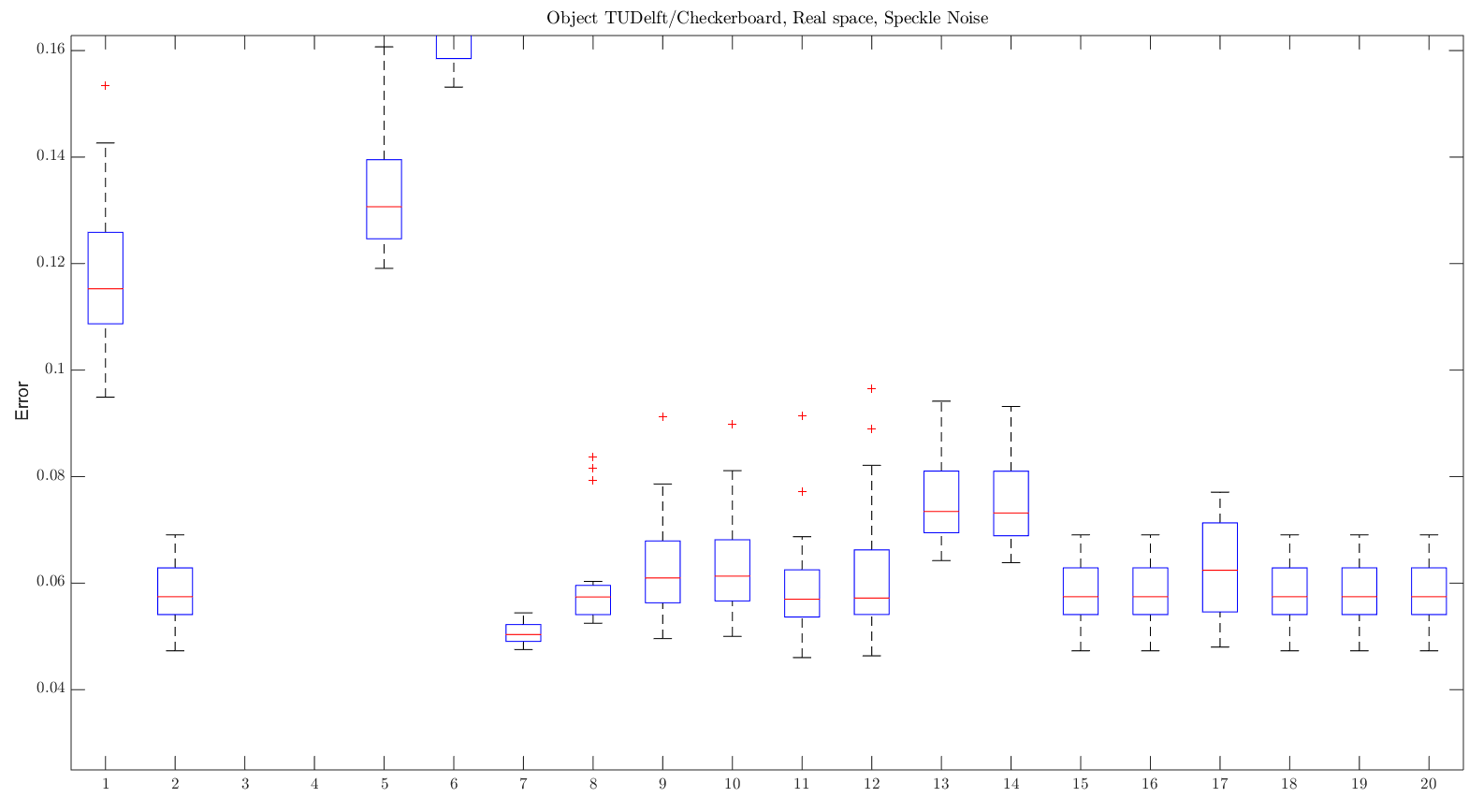}}
\caption{Reconstruction errors for the TUDelft/Checkerboard object (Fig. \ref{fig:Delft}). Each time a constant initial guess is chosen, and the statistics for 20 different realizations of noise are shown.}
\label{fig:resultsA}
\end{figure}

\begin{figure}[H]
\centering
\subfloat[Fourier ptychography, Poisson noise]{\includegraphics[width=0.69\textwidth]{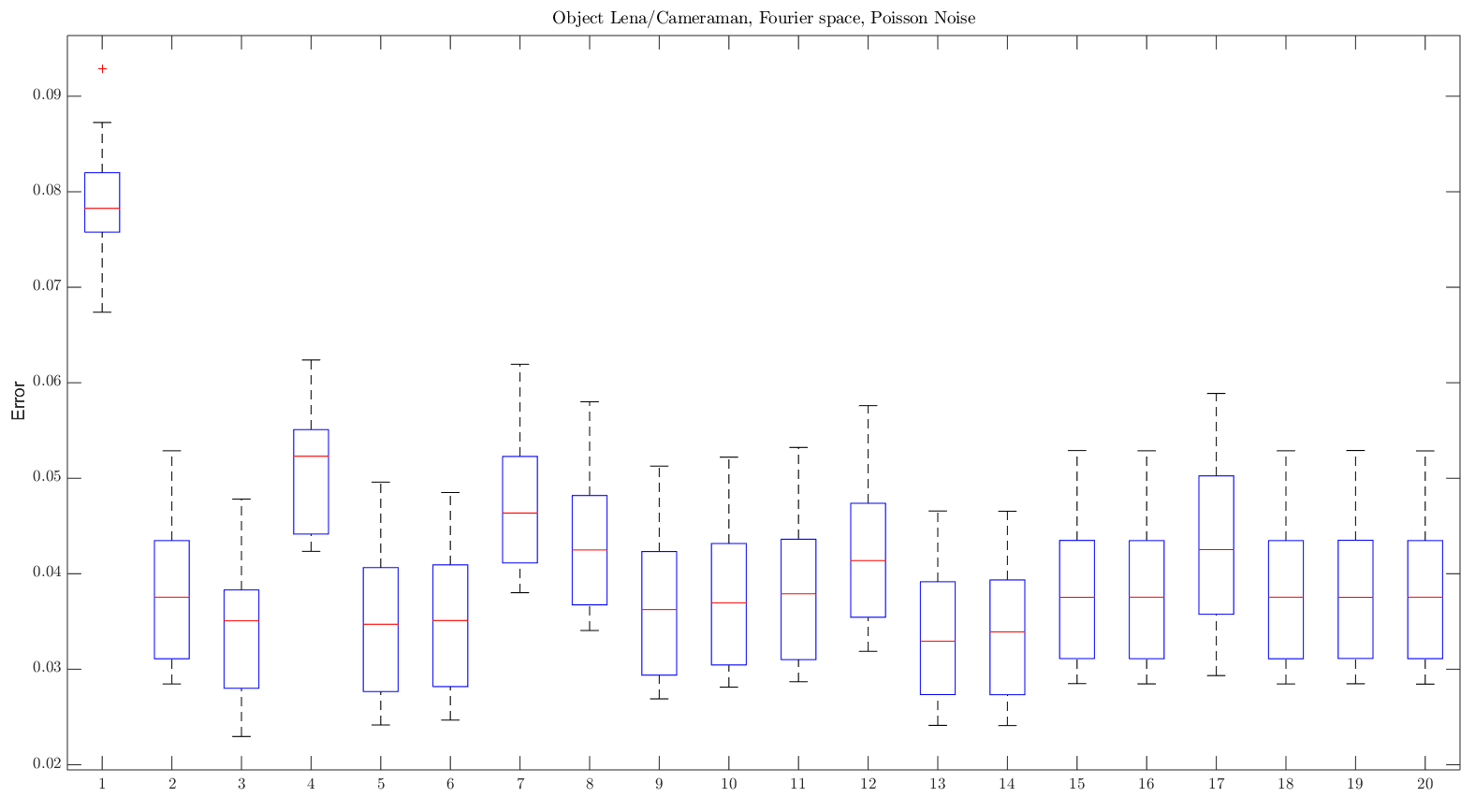}}
\\
\subfloat[Fourier ptychography, Speckle noise]{\includegraphics[width=0.69\textwidth]{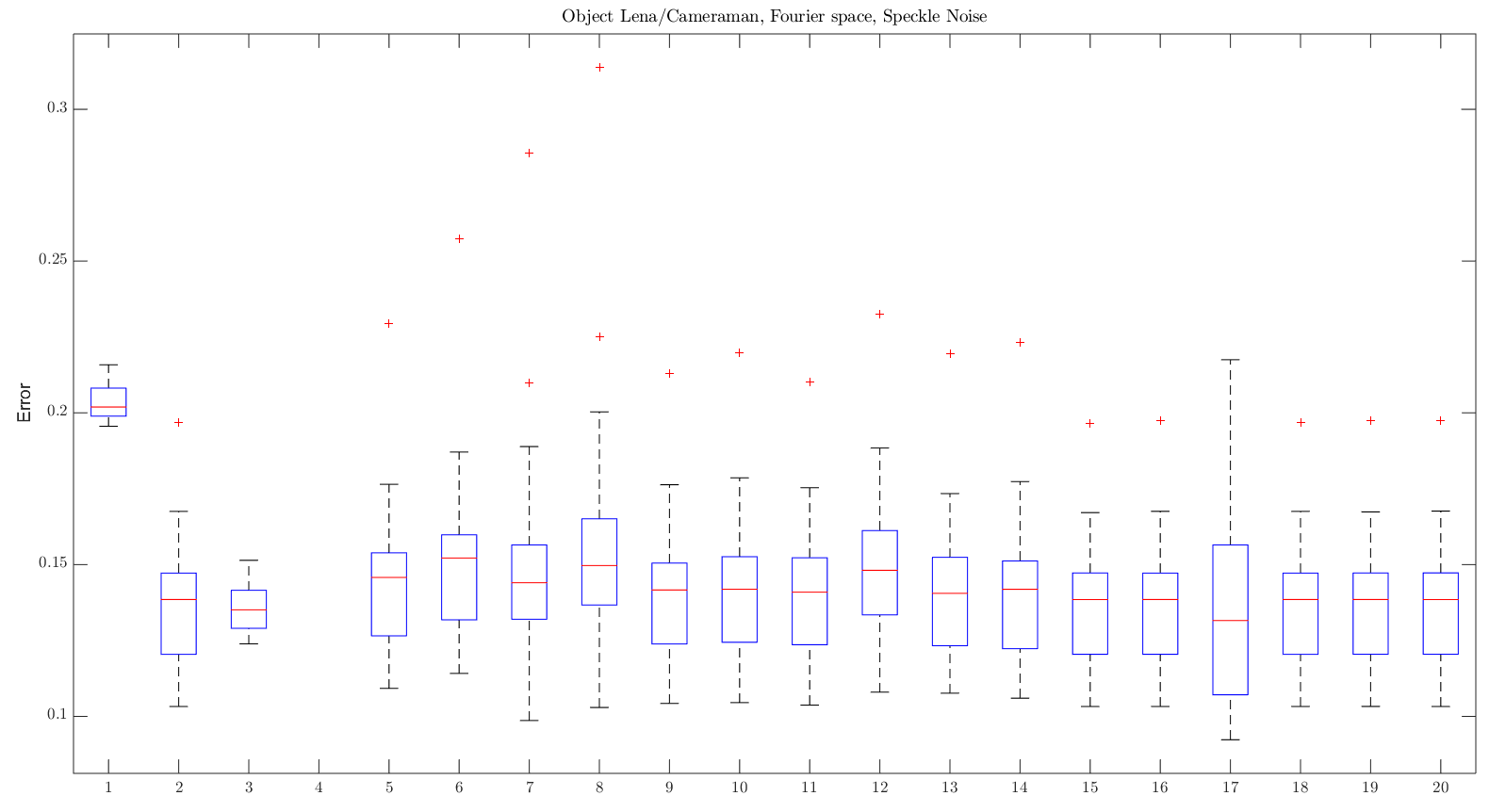}}
\\
\subfloat[Real space ptychography, Poisson noise]{\includegraphics[width=0.69\textwidth]{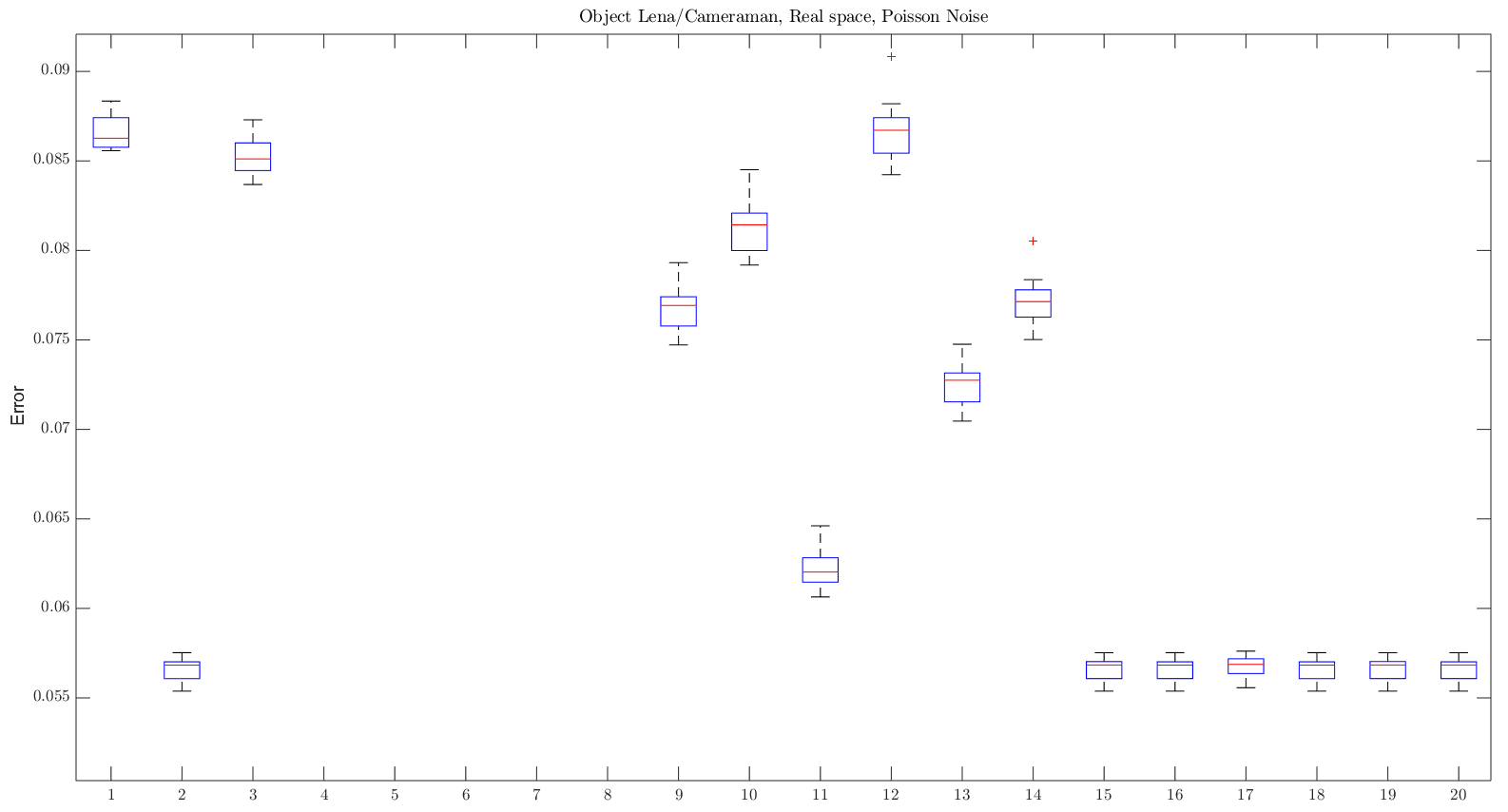}}
\\
\subfloat[Real space ptychography, Speckle noise]{\includegraphics[width=0.69\textwidth]{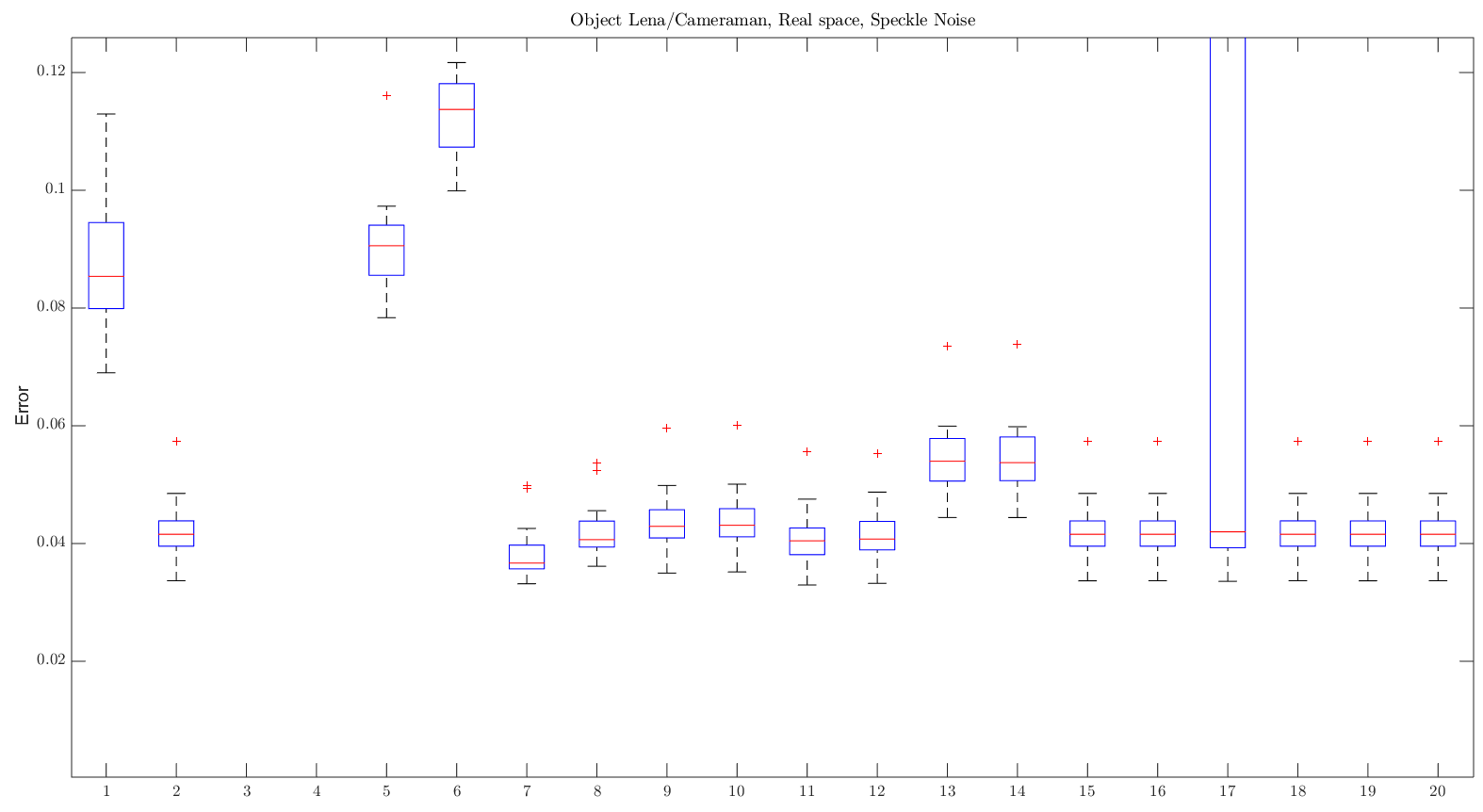}}
\caption{Reconstruction errors for the Lena/Cameraman object (Fig. \ref{fig:Lena}). Each time a constant initial guess is chosen, and the statistics for 20 different realizations of noise are shown.}
\label{fig:resultsB}
\end{figure}

\section{A different proposal: updating the intensity constraints}
\begin{figure}
	\centering
		\includegraphics[width=0.8\textwidth]{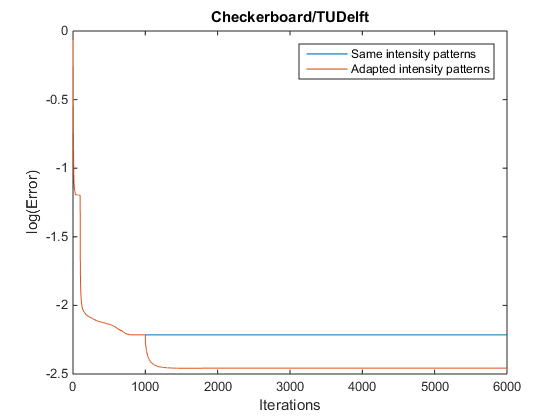}
	\caption{Fourier ptychography, Checkerboard/TUDelft, Photon Count = $10^5$, little oversampling.}
	\label{fig:1}
\end{figure}

\begin{figure}
	\centering
		\includegraphics[width=0.8\textwidth]{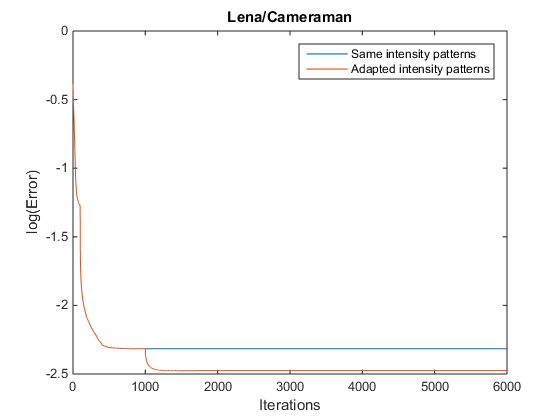}
	\caption{Fourier ptychography, Lena/Cameraman, Photon Count = $10^5$, little oversampling.}
	\label{fig:2}
\end{figure}

\begin{figure}
	\centering
		\includegraphics[width=0.8\textwidth]{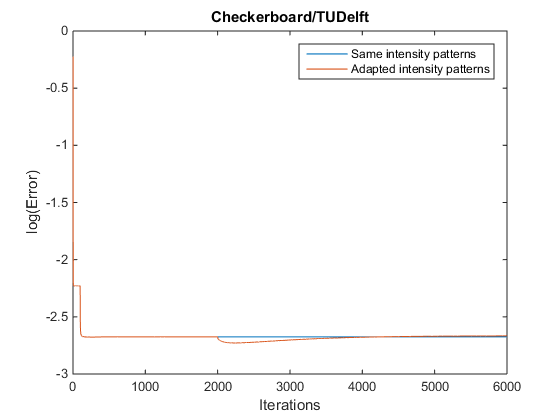}
	\caption{Real space ptychography, Checkerboard/TUDelft, Photon Count = $10^{3.5}$, little oversampling.}
	\label{fig:3}
\end{figure}

\begin{figure}
	\centering
		\includegraphics[width=0.8\textwidth]{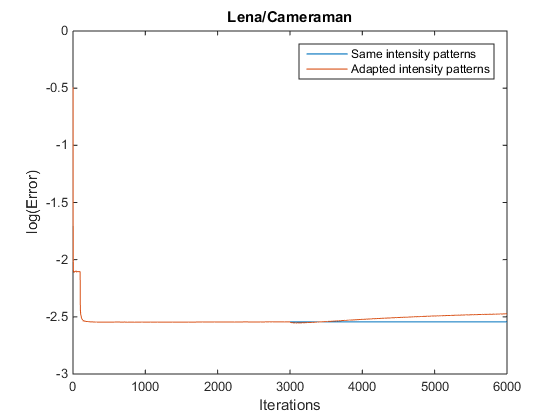}
	\caption{Real space ptychography, Lena/Cameraman, Photon Count = $10^{3.5}$, little oversampling.}
	\label{fig:4}
\end{figure}

\begin{figure}[H]
	\centering
		\includegraphics[width=0.8\textwidth]{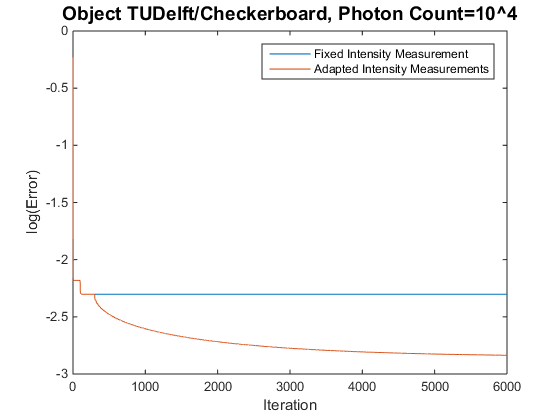}
	\caption{Real space ptychography, Checkerboard/TUDelft, Photon Count = $10^{4}$, oversampling factor = 5.}
	\label{fig:Result1A}
\end{figure}

\begin{figure}[H]
	\centering
		\includegraphics[width=0.8\textwidth]{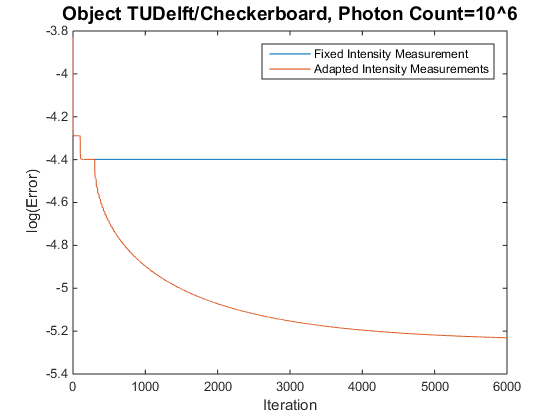}
	\caption{Real space ptychography, Checkerboard/TUDelft, Photon Count = $10^{6}$, oversampling factor = 5.}
	\label{fig:Result2A}
\end{figure}

\begin{figure}[H]
	\centering
		\includegraphics[width=0.8\textwidth]{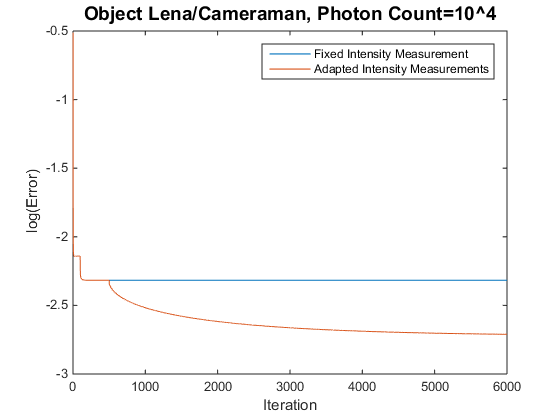}
	\caption{Real space ptychography, Lena/Cameraman, Photon Count = $10^{4}$, oversampling factor = 5.}
	\label{fig:Result1B}
\end{figure}

\begin{figure}[H]
	\centering
		\includegraphics[width=0.8\textwidth]{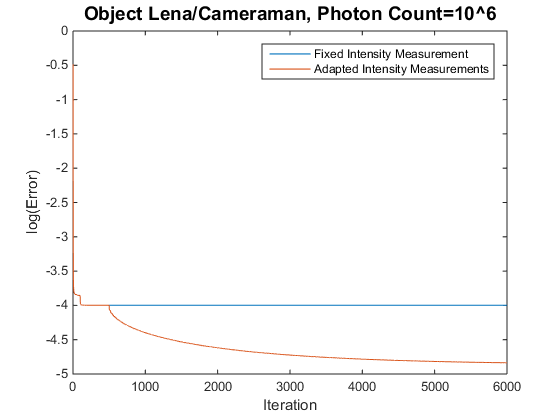}
	\caption{Real space ptychography, Lena/Cameraman, Photon Count = $10^{6}$, oversampling factor = 5.}
	\label{fig:Result2B}
\end{figure}

\begin{figure}[H]
	\centering
		\includegraphics[width=0.8\textwidth]{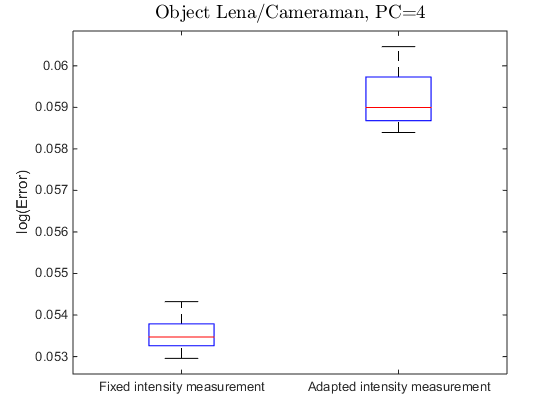}
		\caption{Real space ptychography, Lena/Cameraman, Photon Count = $10^{4}$, little oversampling.}
	\label{fig:Result1Ano}
\end{figure}

\begin{figure}[H]
	\centering
		\includegraphics[width=0.8\textwidth]{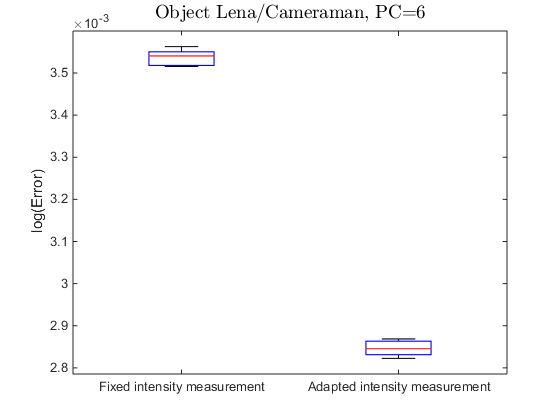}
		\caption{Real space ptychography, Lena/Cameraman, Photon Count = $10^{6}$, little oversampling.}
	\label{fig:Result2Ano}
\end{figure}

\begin{figure}[H]
	\centering
		\includegraphics[width=0.8\textwidth]{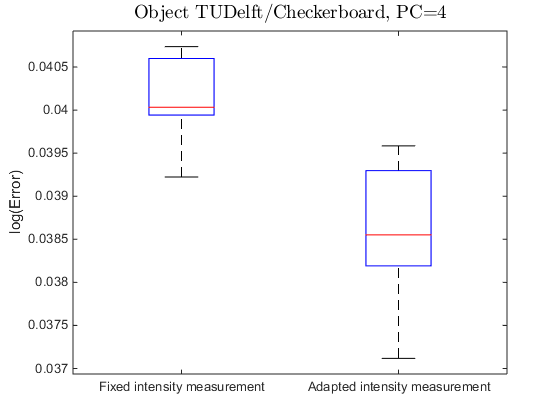}
		\caption{Real space ptychography, Checkerboard/TUDelft, Photon Count = $10^{4}$, little oversampling.}
	\label{fig:Result1Bno}
\end{figure}

\begin{figure}[H]
	\centering
		\includegraphics[width=0.8\textwidth]{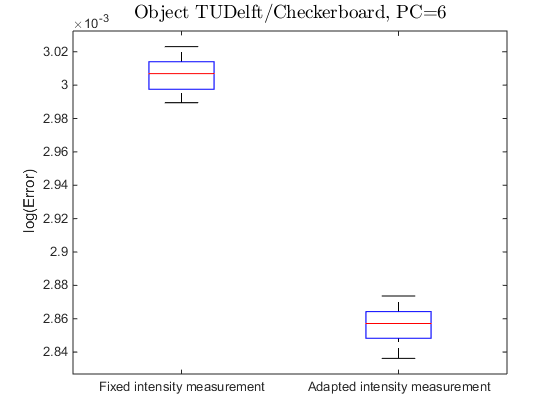}
		\caption{Real space ptychography, Checkerboard/TUDelft, Photon Count = $10^{6}$, little oversampling.}
	\label{fig:Result2Bno}
\end{figure}

Let us look at the problem in the following way. We have reconstructed intensity patterns $z$ which are calculated from the estimated object, we have the measured (noisy) intensity patterns $y$, and there are the (unknown) noise-free intensity patterns $m$. The pixels are still denoted with $u$. The cost functional
\begin{equation}
L=\sum_u (\sqrt{z}-\sqrt{m})^2
\end{equation}
would be ideal, but unknown because we don't know $m$. We do know an approximation of $m$, namely $y$. However, this approximation can be improved by using prior information about the probe positions and the factorization of the exit waves into probe and object. In other words, if we minimize
\begin{equation}
L=\sum_u (\sqrt{z}-\sqrt{y})^2,
\end{equation}
then the resulting estimate should give a better estimate for $m$. Let's define $\tilde{m}$ as our estimated intensity pattern, and $z_0$ as the intensity pattern obtained by minimizing a cost functional $L$. Then we can suggest the following procedure: minimize 
\begin{equation}
L=\sum_u (\sqrt{z}-\sqrt{\tilde{m}_n})^2.
\end{equation}
Use the resulting $z_0$ to define
\begin{equation}
\tilde{m}_{n+1}=(1-\mu) \tilde{m}_{n}+ \mu z_0,
\end{equation}
where $\mu$ is a small step size. Repeat this procedure. Use $\tilde{m}_0=y$. The method of updating the intensity constraints was also applied successfully in \cite{Konijnenberg2016} to make the combination of HIO with PIE more robust against noise.
\\
\\
From simulation results, it seems that for Fourier ptychography adapting the intensity constraints is beneficial (see Figs. \ref{fig:1}, \ref{fig:2}, \ref{fig:3}, \ref{fig:4}). However, for real space ptychography it seems that only if the intensity measurements are well oversampled (by a factor of 5 in these tests), adapting the intensity measurements seems to be beneficial (see Figs. \ref{fig:Result1A}-\ref{fig:Result2B}). If the intensity measurements are barely oversampled, then adapting the intensity measurements seems beneficial when noise levels are not too high (see Figs. \ref{fig:Result1Ano}-\ref{fig:Result2Bno}). Seeing that the oversampling rate affects the performance, it may be required to compare this method to other methods such as binning (in which groups of pixels are lumped together to one pixel). Note though that when the pixel size is finite (and rather large), we don't actually sample the intensity pattern at distinct points, but rather integrate over finite areas. How to correct for this is explained in \cite{Song2007}.

\section{Conclusion}
We have presented an overview of different update schemes for the ptychographic reconstruction algorithm, designed to improve its robustness against noise. Several different viewpoints have been discussed, and a few new ones have been introduced. Even though the reasoning behind them seems plausible, simulation results seem to indicate that its validity is questionable, and that the only consistently reliable way to improve the reconstruction result in the presence of noise is to reduce the step size of the update function as suggested by \cite{Zuo2016}. However, if we know some general characteristics of the object, then depending on the mode of ptychography and the noise model it is possible to find update schemes that reduce the reconstruction error. In addition, a noise-robust ptychographic reconstruction method has been proposed that relies on adapting the intensity constraints. Simulations have shown that this scheme gives improved results when the intensity measurements are sufficiently oversampled, or when the noise level is not too high. However, these results need to be compared more carefully to other noise-reducing methods such as binning. Hopefully, this research advances the understanding of the effect of different update functions in the presence of noise, and perhaps inspires more detailed and more careful research on the topic. 

\bibliographystyle{unsrt}
\bibliography{MyCollection}

\begin{thebibliography}{10}

\bibitem{gerchberg1972practical}
R.~W. Gerchberg and W.~O. Saxton.
\newblock {A practical algorithm for the determination of phase from image and
  diffraction plane pictures}.
\newblock {\em Optik}, 35:237, 1972.

\bibitem{fienup1978reconstruction}
J.~R. Fienup.
\newblock {Reconstruction of an object from the modulus of its Fourier
  transform}.
\newblock {\em Optics Letters}, 3(1):27, jul 1978.

\bibitem{rodenburg2004phase}
J.~M. Rodenburg and H.~M.~L. Faulkner.
\newblock {A phase retrieval algorithm for shifting illumination}.
\newblock {\em Appl. Phys. Lett.}, 85(20):4795, 2004.

\bibitem{maiden2009improved}
A.~M. Maiden and J.~M. Rodenburg.
\newblock {An improved ptychographical phase retrieval algorithm for
  diffractive imaging}.
\newblock {\em Ultramicroscopy}, 109(10):1256--1262, sep 2009.

\bibitem{bauschke2002phase}
H.~H. Bauschke, P.~L. Combettes, and D.~R. Luke.
\newblock {Phase retrieval, error reduction algorithm, and Fienup variants: a
  view from convex optimization}.
\newblock {\em Journal of the Optical Society of America A}, 19(7):1334, jul
  2002.

\bibitem{luke2004relaxed}
D.~R. Luke.
\newblock {Relaxed averaged alternating reflections for diffraction imaging}.
\newblock {\em Inverse Problems}, 21(1):37--50, nov 2004.

\bibitem{bauschke2003hybrid}
H.~H. Bauschke, P.~L. Combettes, and D.~R. Luke.
\newblock {Hybrid projection reflection method for phase retrieval}.
\newblock {\em Journal of the Optical Society of America A}, 20(6):1025, jun
  2003.

\bibitem{Fienup1982}
J.~R. Fienup.
\newblock {Phase retrieval algorithms: a comparison}.
\newblock {\em Appl. Opt.}, 21(15):2758, aug 1982.

\bibitem{GuizarSicairos2008}
M.~Guizar-Sicairos and J.~R. Fienup.
\newblock {Phase retrieval with transverse translation diversity: a nonlinear
  optimization approach}.
\newblock {\em Opt. Express}, 16(10):7264, may 2008.

\bibitem{qian2014efficient}
J.~Qian, C.~Yang, A.~Schirotzek, F.~Maia, and S.~Marchesini.
\newblock {Efficient Algorithms for Ptychographic Phase Retrieval}, 2014.

\bibitem{Konijnenberg2016}
A.~P. Konijnenberg, W.~M.~J. Coene, S.~F. Pereira, and H.~P. Urbach.
\newblock {Combining ptychographical algorithms with the Hybrid Input-Output
  ({\{}HIO{\}}) algorithm}.
\newblock {\em Ultramicroscopy}, 171:43--54, dec 2016.

\bibitem{Yeh2015}
L.~Yeh, J.~Dong, J.~Zhong, L.~Tian, M.~Chen, G.~Tang, M.~Soltanolkotabi, and
  L.~Waller.
\newblock {Experimental robustness of Fourier ptychography phase retrieval
  algorithms}.
\newblock {\em Optics Express}, 23(26):33214, dec 2015.

\bibitem{Thibault2012}
P.~Thibault and M.~Guizar-Sicairos.
\newblock {Maximum-likelihood refinement for coherent diffractive imaging}.
\newblock {\em New Journal of Physics}, 14(6):63004, jun 2012.

\bibitem{Bian2016}
L.~Bian, J.~Suo, J.~Chung, X.~Ou, C.~Yang, F.~Chen, and Q.~Dai.
\newblock {Fourier ptychographic reconstruction using Poisson maximum
  likelihood and truncated Wirtinger gradient}.
\newblock {\em Scientific Reports}, 6:27384, jun 2016.

\bibitem{Godard2012}
P.~Godard, M.~Allain, V.~Chamard, and J.~Rodenburg.
\newblock {Noise models for low counting rate coherent diffraction imaging}.
\newblock {\em Opt. Express}, 20(23):25914, nov 2012.

\bibitem{Marchesini2016}
S.~Marchesini, H.~Krishnan, B.~J. Daurer, D.~A. Shapiro, T.~Perciano, J.~A.
  Sethian, and F.~R. N.~C. Maia.
\newblock {SHARP: a distributed GPU-based ptychographic solver}.
\newblock {\em Journal of Applied Crystallography}, 49(4):1245--1252, jul 2016.

\bibitem{Zhang2017}
Y.~Zhang, P.~Song, and Q.~Dai.
\newblock {Fourier ptychographic microscopy using a generalized Anscombe
  transform approximation of the mixed Poisson-Gaussian likelihood}.
\newblock {\em Optics Express}, 25(1):168, jan 2017.

\bibitem{Anscombe1948}
F.~J. Anscombe.
\newblock {The Transformation of Poisson, Binomial and Negative-Binomial Data}.
\newblock {\em Biometrika}, 35(3/4):246, dec 1948.

\bibitem{Zuo2016}
C.~Zuo, J.~Sun, and Q.~Chen.
\newblock {Adaptive step-size strategy for noise-robust Fourier ptychographic
  microscopy}.
\newblock {\em Optics Express}, 24(18):20724, aug 2016.

\bibitem{Soulez2016}
F.~Soulez, E.~Thi{\'{e}}baut, A.~Schutz, A.~Ferrari, F.~Courbin, and M.~Unser.
\newblock {Proximity operators for phase retrieval}.
\newblock {\em Applied Optics}, 55(26):7412, sep 2016.

\bibitem{Zheng2013}
G.~Zheng, R.~Horstmeyer, and C.~Yang.
\newblock {Wide-field, high-resolution Fourier ptychographic microscopy}.
\newblock {\em Nature Photonics}, 7(9):739--745, jul 2013.

\bibitem{Goodman1976}
J.~W. Goodman.
\newblock {Some fundamental properties of speckle}.
\newblock {\em Journal of the Optical Society of America}, 66(11):1145, nov
  1976.

\bibitem{Song2007}
C.~Song, D.~Ramunno-Johnson, Y.~Nishino, Y.~Kohmura, T.~Ishikawa, C.~Chen,
  T.~Lee, and J.~Miao.
\newblock {Phase retrieval from exactly oversampled diffraction intensity
  through deconvolution}.
\newblock {\em Physical Review B}, 75(1), jan 2007.

\end{thebibliography}

\end{document}